# Melting of Spatially Modulated Phases in La-doped BiFeO$_3$ at Surfaces and Surface-Domain Wall Junctions

*Anna N. Morozovska[1,2], Eugene A. Eliseev[3], Deyang Chen[4], Vladislav Shvetz[2], Christopher T. Nelson[5], and Sergei V. Kalinin[5\**]*

[1] *Institute of Physics, National Academy of Sciences of Ukraine, 46, pr. Nauky, 03028 Kyiv, Ukraine*

[2] *Taras Shevchenko Kyiv National University, Physics Faculty, Kyiv, Ukraine*

[3] *Institute for Problems of Materials Science, National Academy of Sciences of Ukraine, Krjijanovskogo 3, 03142 Kyiv, Ukraine*

[4] *Institute for Advanced Materials and Guangdong Provincial Key Laboratory of Optical Information Materials and Technology, South China Academy of Optoelectronics, South China Normal University, Guangzhou 510006, China*

[5] *The Center for Nanophase Materials Sciences, Oak Ridge National Laboratory, Oak Ridge, TN 37831*

## Abstract

The interplay between the surface and domain wall phenomena in multiferroic La$_x$Bi$_{1-x}$FeO$_3$ in the vicinity of morphotropic phase transition is explored on the atomic level. Scanning Transmission Electron Microscopy (STEM) has enabled mapping of atomic structures of the material with picometer-level precision, providing direct insight into the spatial distribution of the order parameters in this material and their behavior at surfaces and interfaces. Here, we use the thermodynamic Landau-Ginzburg-Devonshire (LGD) approach to explain the emergence of spatially modulated phases (SMP) in La$_{0.22}$Bi$_{0.78}$FeO$_3$ films, and establish that the change of polarization gradient coefficients caused by La-doping is the primary driving mechanisms. The suppression, or "melting", of the SMP in the vicinity of the domain wall surface junction is observed experimentally and simulated in the framework of LGD theory. The melting originated from the system tendency to minimize electrostatic energy caused by long-range stray electric fields outside the film and related depolarization effects inside it. The observed behavior provides insight to the origin of surface and interface behaviors in multiferroics.

---

[*] Corresponding author, e-mail: sergei2@ornl.gov

## I. INTRODUCTION

Multiferroic materials with coupled ferroelectric (**FE**) and ferromagnetic or antiferromagnetic long-range ordering remain at the forefront of modern materials science research. This interest is due both to the broad gamut of current and potential applications and the continuous interest for fundamental physics studies [1, 2, 3, 4, 5]. In particular, applications such as ferroelectric tunneling barriers, light-assisted ferroic dynamics, spin-driven effects, and magnetoelectric switching for memory applications are coming to the forefront of research [6, 7].

Among the materials systems for these applications, particularly of interest are bulk and nanosized multiferroic $BiFeO_3$ (**BFO**) and its solid solutions [8, 9, 10, 11, 12, 13,]. Advances in applications necessitate fundamental understanding of polarization and antiferrodistortive (**AFD**) order parameter dynamics in this material, which in turn necessitates both the study of mesoscale phenomena such as polarization switching, topological defects, and atomic scale phenomena at surfaces and interfaces [14, 15, 16, 17]. On the nanometer scale, the breakthrough in understanding the functional properties of ferroelectrics and multiferroics have been achieved via Scanning Probe Microscopy studies, such as Piezoresponse Force Microscopy (**PFM**) [18, 19] and Scanning Transmission Electron Microscopy [20, 21]. For BFO, these have been used to reveal complex nanoscale evolution of domain structure in thin films [22, 23, 24, 25, 26, 27], including vortices and vertices [28, 29, 30, 31]. Domain walls in BFO were found to exhibit unusual electrophysical properties, such as conduction and enhanced magnetotransport [19, 20, 29, 32, 33, 34].

However, while for pure BFO these studies provide a high veracity insight into polarization dynamics and domain structures, the situation becomes more complex for the rare-earth doped BFO. In this case, in addition to the pure-phase AFD-FE ordering and canted antiferromagnetic subsystem, additional symmetry lowering, spatial modulation, and order parameters can emerge [35, 36, 37, 38, 39, 40, 41]. The polar order parameters predominantly manifest in the A-site sublattice allowing a four sub-lattices model (FSM) [35] to be developed for the analytical description of the corresponding A-cation displacements in $La_xBi_{1-x}FeO_3$ (BFO:La) polymorphs. However, FSM itself cannot provide a link between "additional order parameters" - displacements $A_i$ of La/Bi cations and "intrinsic" long-range FE polarization P and AFD oxygen octahedron tilt Φ. The interplay between $A_i$, P and Φ rules the phenomena taking place at the domain walls, surfaces and interfaces of ortho-ferrites, where the role of the long-range electrostatic and elastic fields conditioned by the spatial confinement can be very significant, if not crucial.

This becomes particularly important in the vicinity of the morphotropic phase boundaries (**MPB**) and associated critical points. There, the nature of the order parameter *per se* becomes somewhat undefined. Similarly, even in the pure BFO, the nature of the boundary terms defining order parameter behavior in the vicinity of surfaces, interfaces, and topological defects is generally unknown [29, 33, 35]. While Landau-Ginzburg-Devonshire (**LGD**) descriptions allow high-veracity description of polarization dynamics on multiple length scales [11, 12, 17], the corresponding coupling terms cannot be determined from mesoscopic models and necessitate atomistic studies.

Scanning Transmission Electron Microscopy (**STEM**) has enabled mapping of atomic structures of solids with precision up to several pm, providing insight to the physics of ferroic phenomena. A number of studies of polarization profiles in the vicinity of the interfaces and domain walls have been reported, and corresponding constants have been extracted. STEM studies have allowed elucidating the enigmatic behavior of spatially modulated phases (**SMP**) and **MPBs** in Rare Earth-doped BFO [39 - 41]. Atomic-scale structure of SMPs at the ferroelectric–antiferroelectric MPB has been visualized by STEM in RE-BFO thin films [39, 40]. Atomically-resolved STEM mapping of structural distortions at multiferroic domains walls revealed that the coexistence of rhombohedral (R-) and orthorhombic (O-) phases in ultrathin BFO films can be driven by interfacial oxygen octahedral coupling [42, 43]. However, the clear physical understanding what are primary order parameters in RE-BFO and how they behave close to surfaces and interfaces is missing [35].

Here, we study the nature of SMP in the La-doped $BiFeO_3$, and explore their behavior in the vicinity of surfaces and domain wall. We develop LGD theory for these phases and argue that the changes of polarization gradient coefficients caused by La-doping is a primary driving force for the formation of modulated phases. The melting of the spatial modulation in the vicinity of twin AFD domain walls junction with the film surface is observed and quantified using FSM model. The SMP melting is explained using LGD-formalism combined with electrostatics. Here, we will show that that SMP melting yields the information on the role of long-range stray electric fields and related depolarization effects in the vicinity of the domain-wall surface junction.

## II. STEM STUDIES OF SPATIALLY MODULATED PHASES OF DOPED BFO

In the bulk rhombohedral R3c phase, bismuth ferrite BiFeO$_3$ (BFO) is a multiferroic material [44, 45] with a large FE polarization, AFD oxygen octahedral rotations, antiferromagnetic order, and long range ferromagnetic order coexisting up to room and elevated temperatures [46, 47]. Bulk BFO exhibits AFD long-range order at temperatures below 1200 K; it is FE with a large spontaneous polarization below 1100 K and is AFM below Neel temperature $T_N \approx 650$ K [48]. Similar to other ferroelectrics, the behavior of the structural order parameter at the domain walls of BFO determines their structure and energy [49].

On doping with La and other rare earth elements, BFO undergoes transition to non-rhombohedral non-ferroelectric phase. Correspondingly, on the phase diagram these symmetry-incompatible phases are joined by the morphotropic phase boundaries. The La doping concentration of 22% used in this work lies near the MPB, exhibiting coexisting ferroelectric R- and antiferroelectric O- phases at room temperature. These phases exhibit distinctive large polar distortions of the La/Bi A-site from pseudocubic positions. In the FE R- phase, displacements are cooperative along the <111> polarization axis. In the antiferroelectric O- phase, modulated displacements occur on alternating pairs of [101]$_{pseudocubic}$ planes [41]. The two phases are readily distinguished by this A-site behavior, this sublattice exhibiting very high signal to noise in HAADF-STEM imaging due to the strong high angle scattering from the large Z cations.

Here, La$_{0.22}$Bi$_{0.78}$FeO$_3$ thin films were fabricated on STO buffered Si substrates with SrRuO$_3$ as the bottom electrode using pulsed laser deposition. The La concentration of the films was experimentally measured to be 22% from Rutherford backscattering spectroscopy. A colorized atomic scale [001] A-site polar displacement map is shown in **Fig.1** derived from HAADF-STEM. Displacements correspond to atom positions relative to their centrosymmetric positions as defined by the neighbor B-site cations. "Striped" regions represent two spatially modulated regions, separated by a 90-degree twin wall. The near surface atomic layers clearly exhibit a suppressed spatial modulation, further called "melted" layer. The O- wall being relatively sharp in the depth of the film, broadens and also exhibits damped modulations approaching the surface. The effect is further called as domain wall "melting".

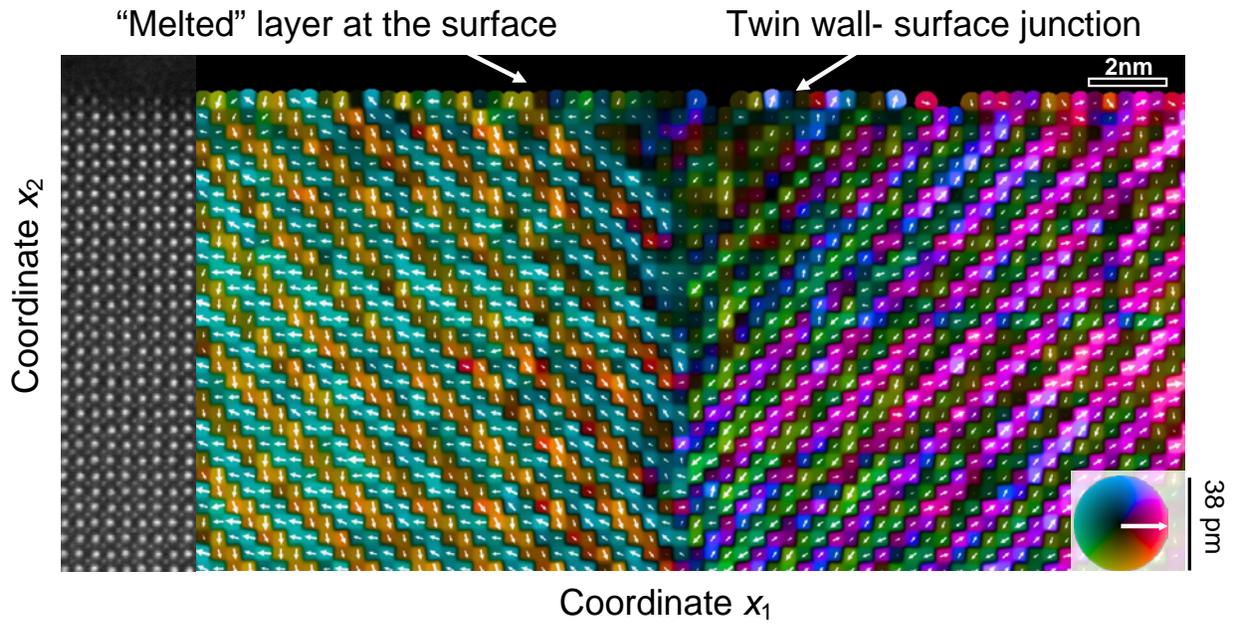

**FIGURE 1.** Colorized map of A-site polar displacements from atomic-resolution HAADF-STEM. The two striped regions represent 90° twin domains of SMP. The twin boundary is sharp in the bulk of the BFO:La film and "melts" approaching the free surface.

To get insight and quantitative description of the behavior of Bi/La polar displacements and link them with "intrinsic" long-range ferroelectric polarization components, the first step is the use of phenomenological four-site model (FSM) [35]. This model describes the displacements of A-cations in four neighboring unit cells (further denoted as $A_i$, $i=1-4$). Using FSM we can reconstruct the behavior of polarization components, corresponding to the STEM mapping shown in **Fig. 1,** using the expression for polar displacement $\vec{P} = \sum_{i=1}^{4} \frac{\vec{A_i}}{2}$ followed from FSM [35]. Reconstruction results for nonzero polar displacement components, $P_1$ and $P_2$, are shown in **Figs. 2.** Symbols are HR STEM experimental data, which $x_1$- and $x_2$-scales are nm, and $P_1$ and $P_2$ are in pm.

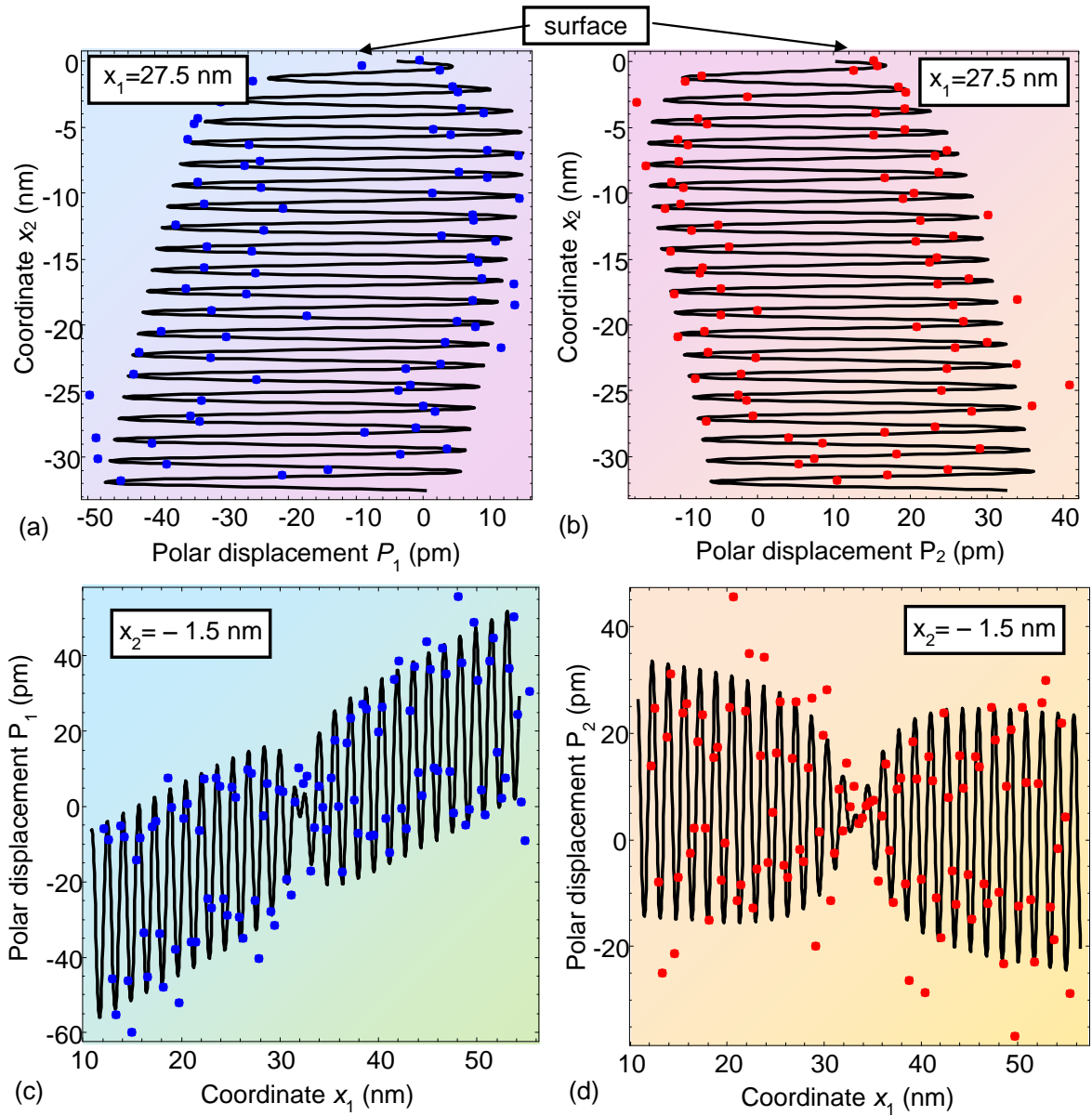

**FIGURE 2.** Profiles of HAADF atomic displacements, corresponding to polarization components $P_1$ (**a, c**) and $P_2$ (**b, d**) in different cross-sections, namely; in planes perpendicular to the film surface near the DW, $x_1 = 27.5$ nm (**a, b**); and in sub-surface planes parallel to the surface, $x_2 = -1.5$ nm (**c, d**). Symbols are experimental data; curves are calculated from Eqs.(1) and (2) with fitting parameters given in **Tables I** and **II**.

It is seen from **Fig. 2a-b** that the amplitude of polarization modulations significantly decreases when approaching the surface clearly indicating the "melting effect", i.e. the reduction of the magnitude of the order parameter on approaching the surface and especially at the junction between the domain wall and the surface. This decrease of the modulation depth at the left-hand sides of **Fig. 2a, 2b** is associated with the reduction of spatial modulation followed by the appearance of homogeneous phase (see **Fig. 1**). The nonzero gradient of the envelope lines from **Fig. 2** could be attributed to the presence of nonzero "offset" and "linear drift" related with the difficulties in separation of polar and nonpolar displacement of ions. Note, that the cross sections of **Fig. 2a** and **2b** are chosen close to domain wall plane, defined by the condition $x_1 = x_{DW}$.

To quantify the melting effect of polarization spatial modulation, the data is approximated by the periodic function shown by solid curves in **Figs. 2.** The chosen functional form allows for periodic modulation with spatially dependent phase and envelope as:

$$P_i(x_1 \approx x_{DW}, x_2) = R_{2i} + u_{2i}x_2 + \alpha_i \sin\left(\frac{2\pi x_2}{t_i} + \psi_{\alpha i}\right) + \beta_i \exp\left(\frac{x_2}{\delta_i}\right) \sin\left(\frac{2\pi x_2}{t_i} + \psi_{\beta i}\right), \quad (1)$$

$$P_i(x_1, x_2 \approx 0) = R_{1i} + u_{1i}x_1 + \frac{\gamma_i(x_1 - x_{DW})}{\sqrt{(x_1 - x_{DW})^2 + \xi_i^2}} \sin\left(\frac{2\pi x_1}{t_i} + \psi_{\gamma i}\right). \quad (2)$$

Here $i$=1, 2 denotes different components of polarization vector and coordinate $x_2 \leq 0$. The fitting parameters $R_{2i}$, $u_{2i}$, $\alpha_i$, $\beta_i$, $\psi_{\alpha i}$, $\psi_{\beta i}$, $R_{1i}$, $u_{1i}$, $\gamma_i$, and $\psi_{\gamma i}$ are summarized in **Table I** and **II**.

**Table I.** Parameters for $x_1$-cut of $P_i(x_1 \approx x_{DW}, x_2)$ fitted by Eq.(1)

| N | $R_{2i}$ (pm) | $u_{2i}$ | $\alpha_i$ (pm) | $\psi_{\alpha i}/\pi$ | $\delta_i$ (nm) | $t_i$ (nm) | $\beta_i$ (pm) | $\psi_{\beta i}/\pi$ |
|---|---|---|---|---|---|---|---|---|
| $i$=1 | −7.24 | 4.32 10$^{-2}$ | −26.3 | 0.228 | 2.41 | 1.61 | 22.7 | 0.362 |
| $i$=2 | 3.43 | −3.70 10$^{-2}$ | −21.2 | 0.240 | 2.41 | 1.61 | 21.5 | 0.452 |

**Table II**. Parameters for $x_2$-cut of $P_i(x_1, x_2 \approx 0)$ fitted by Eq.(2)

| N | $R_{1i}$ (pm) | $u_{1i}$ | $\gamma_i$ (pm) | $\psi_{\gamma i}/\pi$ | $\xi_i$ (nm) | $t_i$ (nm) | $x_{DW}$ (nm) |
|---|---|---|---|---|---|---|---|
| $i$=1 | −46.6 | 13.8 10$^{-2}$ | 21.3 | -0.201 | 1.60 | 1.59 | 32.1 |
| $i$=2 | 12.4 | −2.33 10$^{-2}$ | 20.5 | -0.465 | 4.55 | 1.64 | 33.3 |

The linear combinations $R_{2i} + u_{2i}x_2$ and $R_{1i} + u_{1i}x_1$ in Eqs.(1)-(2) describe the "base" (offset and linear drift ) of the raw HR STEM data and are unrelated to the SMP behavior.

The terms $\beta_i \exp\left(\dfrac{x_2}{\delta_i}\right)$ term in Eq.(1) fits the experimentally observed decay of the spatial modulation amplitude at $x_2 < 0$. The decay corresponds to $\alpha_i < 0$, $\beta_i > 0$, and $\psi_{\alpha i} < \pi/2$, $\psi_{\alpha i} < \pi/2$ in **Table I**. The decay originates from the depolarization field and is exactly responsible for the melting effect. It is seen from **Tables I-II** that the decay length for surface effects $\delta_i$=2.4 nm and characteristic period of oscillations $t_i \approx 1.6$ nm are the same for both components, at that the value 1.6 (nm) is approximately equal to 4 lattice constants (about 0.4 nm) at room temperature. This suggests that melted SMP can be relatively well described by a 2D harmonic modulations $\sim \sin\left(\dfrac{2\pi x_2}{t_i} + \psi_i\right)$ and $\sim \sin\left(\dfrac{2\pi x_1}{t_i} + \psi_i\right)$ in Eqs.(1)-(2).

The power decay prefactor $\dfrac{x_1 - x_{DW}}{\sqrt{(x_1 - x_{DW})^2 + \xi_i^2}}$ in Eq.(2) can point on the existence of long-range stray electric fields, which can cause the DW broadening and SMP melting at the surface, supporting the tendency to minimize the electrostatic energy caused by long-range stray electric fields outside the film and related depolarization effects inside it [50].

However, the observed melting of the SMP in the vicinity of the domain-wall surface junction can be attributed to different mechanisms, which will be explored later using continuous LGD approach.

## III. LANDAU-GINZBURG-DEVONSHIRE FORMALISM

FSM allows the phenomenological description of the melting process. Here, we aim to explore its driving mechanism within LGD approach using thermodynamic functional of BFO:La.

### A. Thermodynamic potential

Continuum medium approaches, such as LGD thermodynamic potential combined with electrostatic equations and elasticity theory, allows self-consistent determination of polarization, structural order, electric and elastic fields in bulk and nanosized ferroics, their surface, interfaces, antiphase boundaries, 180-degree and twin domain walls of arbitrary geometry. LGD functional of AFD-FE multiferroic (such as BFO:La) utilizes Landau-type power expansion, that includes FE and AFD orders, the biquadratic couplings between the order parameters, as well LGD potential is [8, 11, 12, 43]:

$$G_{LGD} = \int_{-\infty}^{\infty} dx_3 \int_{-\infty}^{\infty} dx_1 \int_0^h \left( G_{P+\Phi} + G_{\nabla\Phi} + G_{\nabla P} + G_{P\Phi} + G_{EL} + G_{FL} \right) dx_2 + G_S. \quad (3a)$$

It contains separate contributions of polarization components $P_i$ and components of the pseudo-vector determining the out-of-phase static rotations of the oxygen octahedrons $\Phi_i$,

$$\begin{aligned} G_{P+\Phi} &= a_i(T) P_i^2 + a_{ij} P_i^2 P_j^2 + a_{ijk} P_i^2 P_j^2 P_k^2 - P_i E_i - \frac{\varepsilon_0 \varepsilon_b}{2} E_i^2 \\ &+ b_i(T) \Phi_i^2 + b_{ij} \Phi_i^2 \Phi_j^2 + b_{ijk} \Phi_i^2 \Phi_j^2 \Phi_k^2 \end{aligned} \quad (3b)$$

Here Einstein summation convention is employed over repeated indexes. The coefficients $a_k$ and $b_i$ are temperature dependent, $a_k^{(P)} = \alpha_T \left( T_{qP} \coth(T_{qP}/T) - T_C \right)$ and $b_i = b_T T_{q\Phi} \left( \coth(T_{q\Phi}/T) - \coth(T_{q\Phi}/T_\Phi) \right)$, where $T_C$ is a Curie temperature, $T_\Phi$ is the AFD transition temperature, $T_{qP}$ and $T_{q\Phi}$ are the characteristic Barrett-type temperature related with some "vibrational modes" [51, 52]. $E_i$ are the components of internal electric field related with electrostatic potential $\varphi$ in a standard way $E_i = -\partial\varphi/\partial x_i$. The potential $\varphi$ can be determined from electrostatic equations in a self-consistent manner. Inside the ferroelectric film, the potential satisfies the Poisson equation, $\varepsilon_0 \varepsilon_b \Delta\varphi - \text{div}\vec{P} = 0$, where $\varepsilon_0$ is a universal dielectric constant, $\varepsilon_b$ is the dielectric permittivity of background [53].

The gradient energies are

$$G_{\nabla P} = g_{ijkl} \frac{\partial P_i}{\partial x_k} \frac{\partial P_j}{\partial x_l} + w_{ijkl} \left(\frac{\partial P_i}{\partial x_k}\right)^2 \left(\frac{\partial P_j}{\partial x_l}\right)^2, \quad (3c)$$

$$G_{\nabla\Phi} = v_{ijkl} \frac{\partial \Phi_i}{\partial x_k} \frac{\partial \Phi_j}{\partial x_l} + h_{ijkl} \left(\frac{\partial \Phi_i}{\partial x_k}\right)^2 \left(\frac{\partial \Phi_j}{\partial x_l}\right)^2. \quad (3d)$$

Biquadratic coupling $G_{P\Phi}$, and elastic energy $G_{EL}$:

$$G_{P\Phi} = \zeta_{ijkl} \Phi_i \Phi_j P_k P_l, \quad G_{EL} = -s_{ijkl} \sigma_{ij} \sigma_{kl} - Q_{ijkl} \sigma_{ij} P_k P_l - R_{ijkl} \sigma_{ij} \Phi_k \Phi_l. \quad (3e)$$

The temperature-independent coefficients $\zeta_{ijkl}$ are the components of biquadratic AFD-FE coupling tensor. The elastic energy includes electrostrictive ($\sim Q_{ijkl}$) and rotostrictive ($\sim R_{ijkl}$) contributions. Elastic stress tensor $\sigma_{ij}$ is self-consistently determined from elasticity theory equations. Values $s_{ijkl}$ are the components of elastic compliances tensor.

Flexoelectric-energy is written in the form of bilinear and/or nonlinear Lifshitz invariants [29, 39, 54],

$$G_{FL} = \frac{F_{ijkl}}{2}\left(\sigma_{ij}\frac{\partial P_k}{\partial x_l} - P_k\frac{\partial \sigma_{ij}}{\partial x_l}\right) + \Psi_{ijkl}\left(\Phi_i\Phi_j\frac{\partial P_k}{\partial x_l} - P_k\frac{\partial(\Phi_i\Phi_j)}{\partial x_l}\right), \tag{3f}$$

Here $F_{ijkl}$ are flexoelectric and $\Psi_{ijkl}$ are flexo-antiferrodistortive coupling constants.

The surface energy $G_S$ has the form:

$$G_S = \int_{-\infty}^{\infty} dx_3 \int_{-\infty}^{\infty}\left(\frac{b_i^{(S)}}{2}\Phi_i^2 + \frac{a_i^{(S)}}{2}P_i^2\right)dx_1 \tag{3g}$$

## B. Spatially modulated phases origin

LGD theory allows for several mechanisms of the SMP appearance in AFD-FE ferroics, based on the gradient couplings between the **P**, **Φ** and $\sigma_{ij}$. On one hand, the increase of flexoelectric- and/or flexo-antiferrodistortive energy, written in the form of bilinear and/or nonlinear Lifshitz invariants (3f), can lead to the SMP under the increase of flexoelectric $F_{ijkl}$ and/or flexo-antiferrodistortive $\Psi_{ijkl}$ coupling constants [29, 39, 54]. On the other hand, intrinsic structural instabilities of domain walls driven by polarization gradient couplings (decrease or sign change of the gradient coefficients $g_{ijkl}$ or/and $v_{ijkl}$) was predicted in BFO using LGD theory combined with electrostatics [55]; and then the gradient-driven morphological phase transition at the conductive 180-, 109- and 71- degree domain walls of strained BFO films has been revealed by HR STEM [56]. However the possible driving forces of SMP in thin films of BFO:La was not studied.

To explain the observation of SMP in thin La:BFO film, here we simulated the conditions of the SMP phase appearance at the wall surface junction using LGD-formalism combined with electrostatics and elasticity theory.

### C. Finite element modeling (FEM) of BFO: La film with SMP

Here, we consider a BFO:La film of thickness $h$ placed in a perfect electric contact with conducting bottom electrode that mechanically clamps the film at the surface $x_2 = 0$ (see **Fig. 3**). The top surface of the film ($x_2 = h$) is mechanically free and partially electrically open, e.g. it can be separated from the top electrode by an ultra-thin gap, or covered with the surface screening charge, appearing due to surface states [57], or electro-chemically active ions [58, 59, 60, 61].

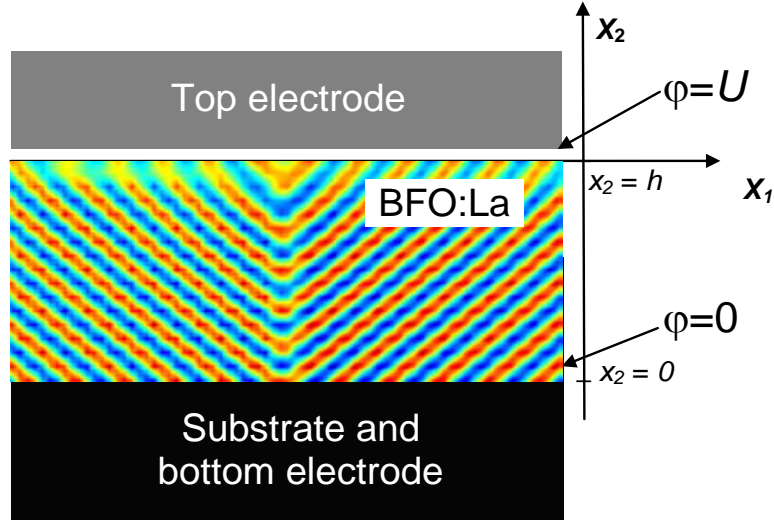

**FIGURE 3**. Considered geometry of the BFO:La film with melted spatial modulation near the surface.

The system of coupled Euler-Lagrange equations allowing for Khalatnikov relaxation of the oxygen tilt and polarization components, $\Phi_i$ and $P_i$, is:

$$\frac{\delta G_{LGD}}{\delta P_i} = -\Gamma_P \frac{\partial P_i}{\partial t}, \qquad \frac{\delta G_{LGD}}{\delta \Phi_i} = -\Gamma_\Phi \frac{\partial \Phi_i}{\partial t}. \tag{4a}$$

Here $i=1, 2, 3$. Material and fitting parameters are listed in **Table III.** Equations (4a) are supplemented by the boundary conditions of zero generalized fluxes at the film boundaries,

$$\left. b^{(S)}\Phi_i + v_{ijkl}\frac{\partial \Phi_j}{\partial x_k}n_l \right|_{x_2=0,h} = 0, \quad \left. a^{(S)}P_i + g_{ijkl}\frac{\partial P_j}{\partial x_k}n_l \right|_{x_2=0,h} = 0. \tag{4b}$$

Surface energy coefficients $b_i^{(S)}$ and $a_i^{(S)}$ in Eqs.(4b) rule the order parameters behavior near the surface as shown in **Fig. S1, Suppl.Mat.,** at that zero values, $\Phi_i = 0$ and $P_i = 0$, at the surface $x_2 = h$ increase the broadening of the domain wall, and the "natural" conditions of zero fluxes,

$$\left. v_{ijkl}\frac{\partial \Phi_j}{\partial x_k}n_l \right|_{x_2=0,h} = 0, \quad \left. g_{ijkl}\frac{\partial P_j}{\partial x_k}n_l \right|_{x_2=0,h} = 0, \text{ decrease the broadening effect.}$$

Elastic problem formulation is based on the modified Hooke's law obtained using the thermodynamic relation $u_{ij} = -\dfrac{\delta G_{ELS}}{\delta \sigma_{kl}}$, where $u_{ij}$ are the components of elastic strain tensor. Mechanical equilibrium conditions are $\partial \sigma_{ij}/\partial x_j = 0$. The film-substrate interface is strained, because misfit strain close to -1% corresponds to BFO/STO pair. Note that a misfit strain can affect strongly on the film polar properties [62].

**Table III.** Parameters used in FEM for BFO and BFO:La

| Parameter | Designation | Numerical values for BFO (BFO:La) | Ref |
|---|---|---|---|
| Effective permittivity | $\varepsilon_{eff} = \Sigma_i \varepsilon_{bi} + \varepsilon_{el}$ | 7 | f.p. |
| dielectric stiffness | $\alpha_T$ ($\times 10^5$ C$^{-2}\cdot$Jm/K) | 9 | [11] |
| Curie temperature for P | $T_C$ (K) | 1300 | [11] |
| Barret temperature for P | $T_{qP}$ (K) | 800 | [11] |
| polar expansion 4$^{th}$ order | $a_{ij}$ ($\times 10^8$ C$^{-4}\cdot$m$^5$J) | $a_{11} = -13.5$, $a_{12} = 5$ | [11] |
| LGD expansion 6$^{th}$ order | $a_{ijk}$ ($\times 10^9$ C$^{-6}\cdot$m$^9$J) | $a_{111} = 11.2$, $a_{112} = -3$, $a_{123} = -6$ | [11] |
| electrostriction | $Q_{ij}$ (C$^{-2}\cdot$m$^4$) | $Q_{11} = 0.054$, $Q_{12} = -0.015$, $Q_{44} = 0.02$ | [12] |
| Stiffness components | $c_{ij}$ ($\times 10^{11}$ Pa) | $c_{11} = 3.02$, $c_{12} = 1.62$, $c_{44} = 0.68$ | [62] |
| polarization gradient coefficients | $g_{ij}$ ($\times 10^{-10}$ C$^{-2}$m$^3$J) | BFO – $g_{11} = 8$, $g_{12} = -0.5$, $g_{44} = 5$<br>BFO:La – $g_{11} = 10$, $g_{12} = -7$, $g_{44} = 5$ | f.p. |
| AFD-FE coupling | $\xi_{ij}$ ($\times 10^{29}$ C$^{-2}\cdot$m$^{-2}$ J/K) | $\xi_{11} = -0.5$, $\xi_{12} = 0.5$, $\xi_{44} = -2.6$ | [11] |
| tilt expansion 2$^{nd}$ order | $b_T$ ($\times 10^{26}\cdot$J/(m$^5$K)) | 4 | [11] |
| Curie temperature for $\Phi$ | $T_\Phi$ (K) | 1440 | [11] |
| Barret temperature for $\Phi$ | $T_{q\Phi}$ (K) | 400 | [11] |
| tilt expansion 4$^{nd}$ order | $b_{ij}$ ($\times 10^{48}$ J/m$^7$) | $b_{11} = -24 + 4.5\,(\coth(300/T) - \coth(3/14))$<br>$b_{12} = 45 - 4.5\,(\coth(300/T) - \coth(1/4))$ | [11] |
| tilt expansion 6$^{nd}$ order | $b_{ijk}$ ($\times 10^{70}$ J/m$^9$) | $b_{111} = 4.5 - 3.4\,(\coth(400/T) - \coth(2/7))$<br>$b_{112} = 3.6 - 0.04\,(\coth(10/T) - \coth(1/130))$<br>$b_{123} = 41 - 43.2\,(\coth(1200/T) - \coth(12/11))$ | [11] |
| tilt gradient coefficients | $v_{ij}$ ($\times 10^{11}$ J/m$^3$) | $v_{11} = 2$, $v_{12} = -1$, $v_{44} = 1$ | f.p. |
| rotostriction | $R_{ij}$ ($\times 10^{18}$ m$^{-2}$) | $R_{11} = -1.32$, $R_{12} = -0.43$, $R_{44} = 8.45$ | [12] |
| Flexoelectric coefficients | $F_{ij}$ ($\times 10^{-11}$ m$^3$/C) | $F_{11} = 2$, $F_{12} = 1$, $F_{44} = 0.5$ | f.p. |
| Surface energy coefficients | $b_i^{(S)}$ and $a_i^{(S)}$ | $b_i^{(S)} \to \infty$ corresponds to $\Phi_i = 0$ on the surface of the film S, and the condition $a_i^{(S)} = 0$ corresponds to $\partial P_i/\partial n = 0$ | f.p. |

*f.p. - fitting parameter

Distributions of nonzero components $\Phi_1$ and $\Phi_2$ AFD order parameter, FE polarization components $P_1$ and $P_2$, elastic stress and electric potential were simulated by FEM near the O- twin-wall surface junction in BFO: La and undoped 109° domain wall in BFO thin films by LGD-approach. We superposed a random seeding at a regular 109° twin-wall and studied the system relaxation to an equilibrium state. Also we imposed natural boundary conditions for polarization components, and regarded that tilts are zero at both surfaces. Results of FEM are shown in **Fig. 4** and **Fig. 5-6**, respectively, where we showed the final equilibrium state.

FEM simulations for pure BFO exhibit expected distributions of tilts and polarization components at AFD-FE wall approaching the surface (see **Fig. 4**). $\Phi_1$ and $P_1$ are of the same sign on opposite sides of the wall (see **Figs. 4a,d**). $\Phi_2$ and $P_2$ have different signs on opposite sides of the wall (see **Figs. 4b,e**). The wall width in the central part of the film is about 2.5 nm, and about 5 nm at the surface. The twice broadening is conditioned by the system tendency to minimize the depolarization field energy [50], resulting into reduction of electrostatic potential values (see **Fig. 4f**). Piezoelectric effect causes the surface deformation (see **Figs. S2-3** in **Suppl. Mat**).

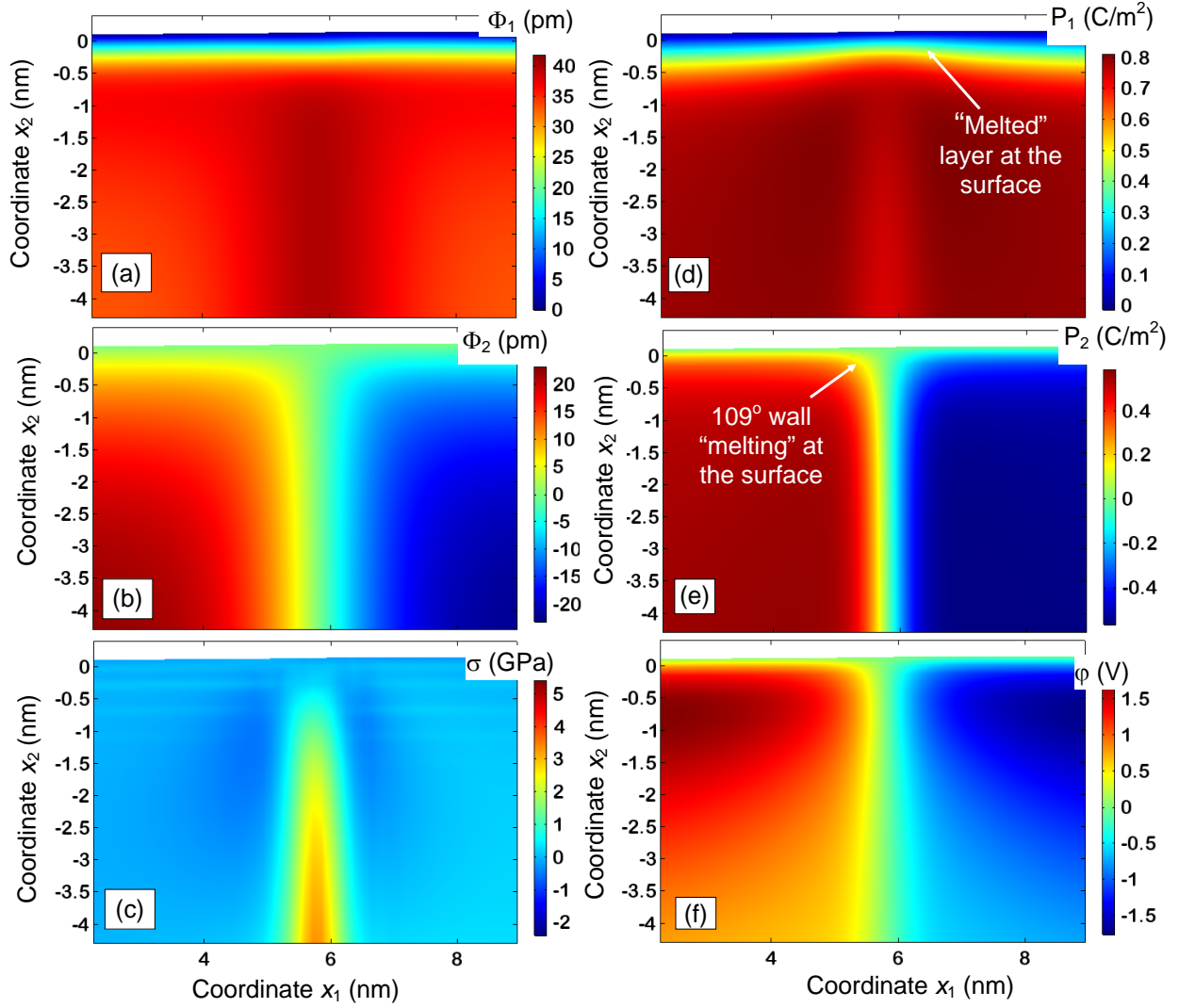

**FIGURE 4**. Distribution of tilts **Φ (a-b**, in pm**),** hydrostatic pressure σ **(c**, in TGa**),** polarization **P (d-e**, in C/m$^2$), and electrostatic potential φ **(f,** in Volts**)** near the twin-wall surface junction in 10-nm BFO film. Here we supposed natural boundary conditions for polarization components and zero tilts at the surfaces. BFO parameters are listed in **Table III,** $T$=293 K**.**

FEM simulations for BFO:La exhibit spatially modulated distributions of tilts and polarization components throughout the entire film (see **Figs. 5-6).** The modulation period (about 0.4 nm) is conditioned by the system tendency to minimize the imbalance between the gradient and depolarization energies. Since the period coincides with a lattice constant, the SMP mimic AFE phase. $\Phi_1$ has the same sign, while $\Phi_2$ has different sign on opposite sides of the wall (compare **Figs. 5a**, and **5b,** and **6a**). Elastic stress is concentrated in the modulation rather than at domain wall (see **Fig. 5c**). With respect to Φ the SMP is rippled, because a nonzero background dominates (see **Figs. 6a,b**). Since $P_1$ and $P_2$ (as well as φ) have different signs on opposite sides of the wall (see **Figs. 5d**-**f**), the SMP is complete with respect to **P**, because any background is absent (see **Figs. 6c,d**). The spatial modulation of polarization "melts" across the AFD-FE wall, and the melting effect becomes more pronounced when the wall approaches the surface (see **Fig. 6**). The origin for the melting effect is the system tendency to reach optimal balance between the gradient-correlation and electrostatic energies.

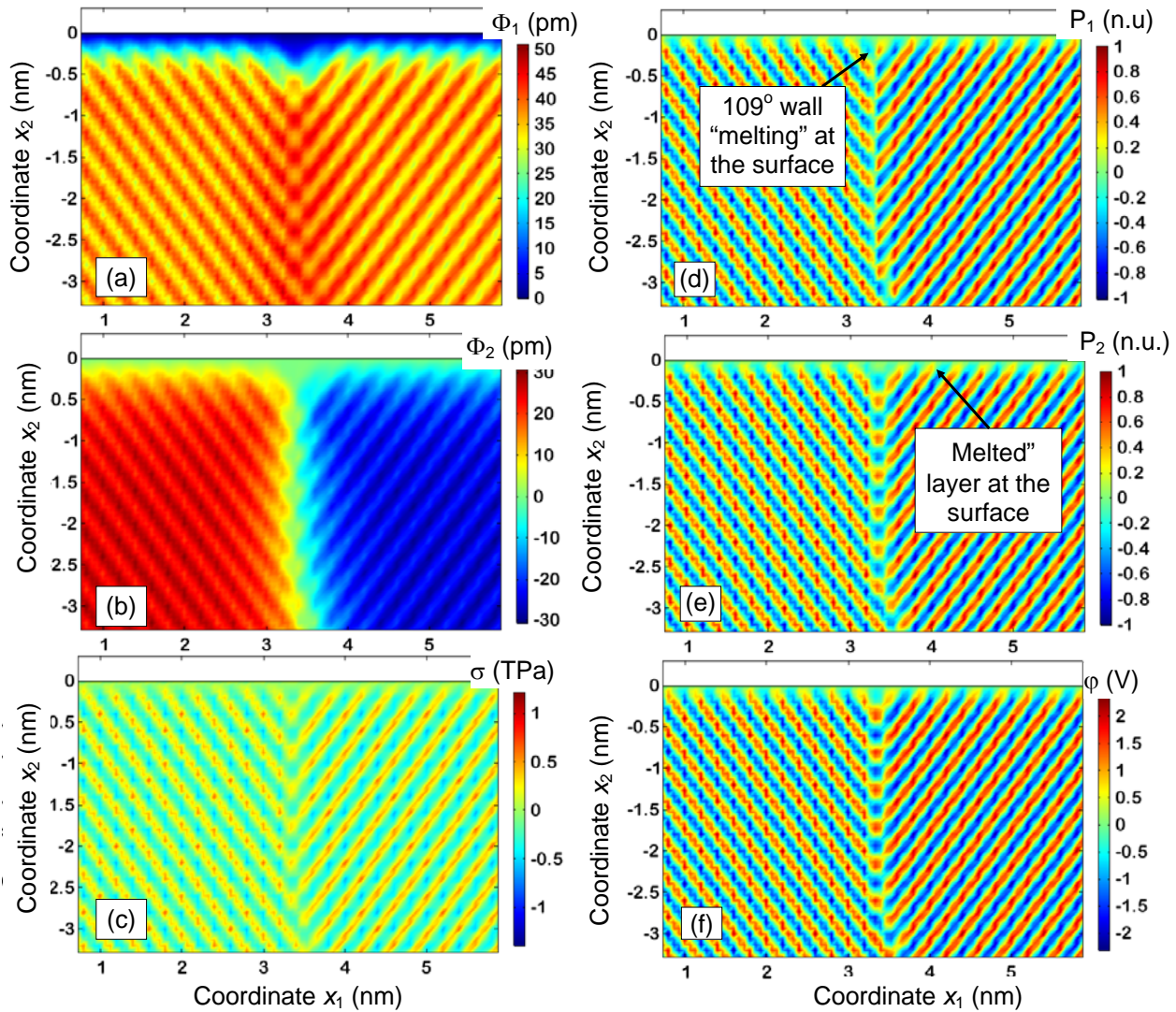

**FIGURE 5**. Distribution of tilts **Φ** (**a-b**, in pm**),** hydrostatic pressure σ (**c**, in TPa**),** polarization **P** (**d-e**, in normalized units), and electrostatic potential φ (**f,** in Volts**)** near the twin-wall surface junction in a 10-nm

BFO:La film. Here we supposed natural boundary conditions for polarization components and zero tilts at the surfaces. BFO:La parameters are listed in **Table III,** $T$=293 K, $h$=10 nm**.**

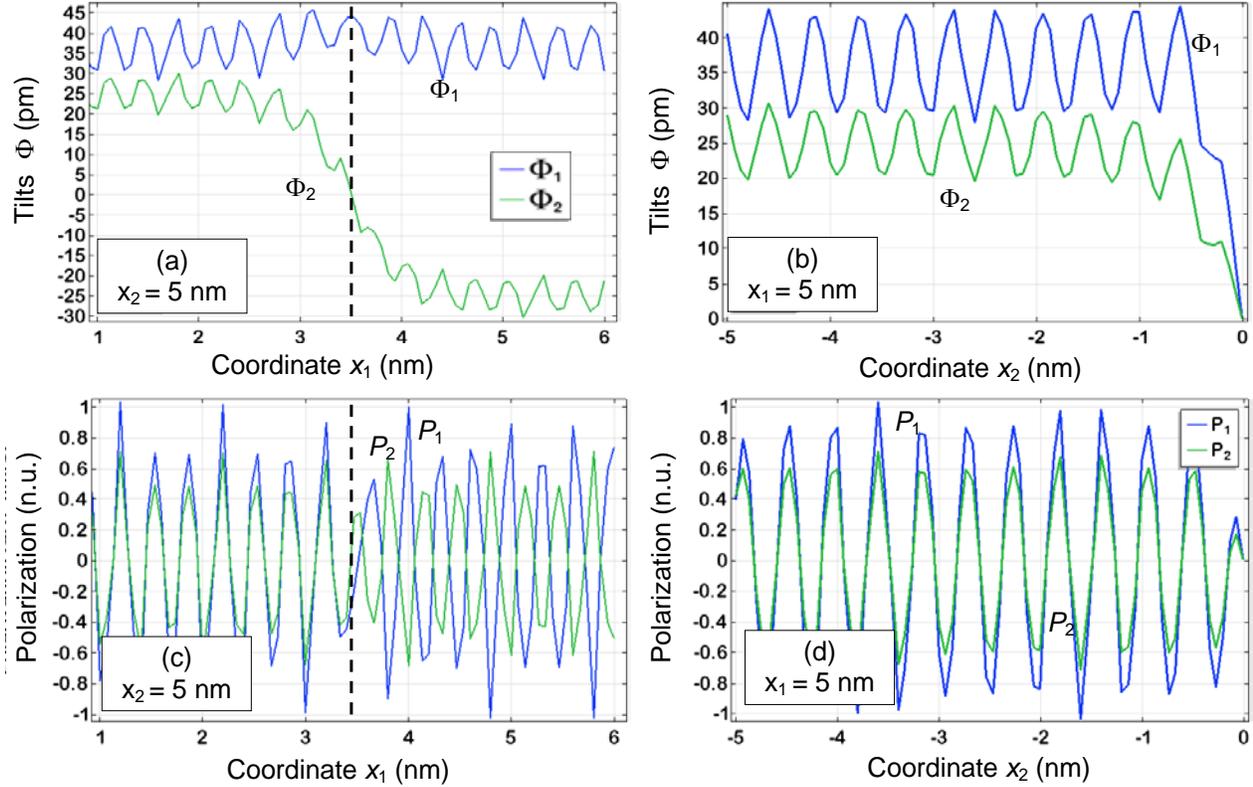

**FIGURE 6**. Profiles of the tilt $\Phi$ (**a-b,** in pm) and polarization **P** (**c-d,** in normalized units) across and along the twin-wall surface junction in a 10-nm BFO:La film. Here we supposed natural boundary conditions for polarization components and zero tilts at the surfaces. BFO:La parameters are listed in **Table III,** $T$=293 K**.**

LGD-based FEM results shown in **Figs. 5** and **6** are in a semi-quantitative agreement with HR STEM data shown in **Figs. 1** and **2**, respectively. Thus the origin of the SMP is the increase of polarization gradient energy in BFO:La film. Note that LGD parameters of BFO;La are different from BFO ones only by the gradient terms $g_{ijkl}$ (see **Table III**)**.** Namely, $g_{11}$=10, $g_{12}$= −7, $g_{44}$=5 in $10^{-10}C^{-2}m^{3}J$ for BFO:La, while $g_{11}$=8, $g_{12}$= −0.5, $g_{44}$=5 in $10^{-10}C^{-2}m^{3}J$ for BFO. The changes of the gradient coefficients (the biggest discrepancy corresponds to $g_{12}$) occur at La addition to BFO are related with the changes of exchange-correlation interaction between neighboring B-cations, since La atoms substitutes Bi atoms in the lattice.

Hence our LGD-based modeling corroborates the statement that the driving forces of polarization spatial modulation melting in the vicinity of the domain-wall surface junction in BFO:La film is the balance of the order parameters gradient energy, and electrostatic energy caused by long-range stray electric fields outside the film, and related depolarization fields inside it.

## IV. CONCLUSION

The interplay between the surface and domain wall phenomena in multiferroic $La_xBi_{1-x}FeO_3$ in the vicinity of morphotropic phase transition is explored using atomically resolved STEM imaging. The formation of periodically modulated phase in the bulk and their melting in the vicinity of the surface and especially surface-domain wall junction is observed and quantified on the atomic level. The thermodynamic LGD approach is used to explain the emergence of SMP in $La_xBi_{1-x}FeO_3$ films, and establish that the primary driving mechanisms for phase formation is the change of polarization gradient coefficients caused by La-doping.

The melting of the SMP in the vicinity of the domain wall surface junction is observed experimentally and simulated in the framework of LGD theory. The melting originated from the system tendency to minimize electrostatic energy caused by long-range stray electric fields outside the film and related depolarization effects inside it. The observed behavior provides insight to the origin of surface and interface behaviors in multiferroics.

**Acknowledgements.** This material is based upon work (S.V.K, C.T.N.) supported by the U.S. Department of Energy, Office of Science, Office of Basic Energy Sciences, and performed in the Center for Nanophase Materials Sciences, supported by the Division of Scientific User Facilities. A portion of FEM was conducted at the Center for Nanophase Materials Sciences, which is a DOE Office of Science User Facility (CNMS Proposal ID: 257). D.C. thanks the financial support from the National Natural Science Foundation of China (Grant Nos. U1832104 and 11704130), the Guangzhou Science and Technology Project (Grant No. 201906010016) and Guangdong Provincial Key Laboratory of Optical Information Materials and Technology (No. 2017B030301007). A.N.M work supported by the National Academy of Sciences of Ukraine and has received funding from the European Union's Horizon 2020 research and innovation programme under the Marie Skłodowska-Curie grant agreement No 778070.

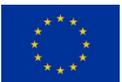

**Authors' contribution.** A.N.M. and E.A.E. proposed the theoretical model and performed calculations. D.C. performed the LBFO film PLD synthesis. C.T.N. performed and analyzed the STEM experiments. S.V.K. generated the research idea, interpreted theoretical and experimental results and wrote the manuscript draft. All co-authors worked on the results discussion and manuscript improvement.

[37] C.-J. Cheng, A.Y. Borisevich, D. Kan, I. Takeuchi, and V. Nagarajan. "Nanoscale structural and chemical properties of antipolar clusters in Sm-doped BiFeO3 ferroelectric epitaxial thin films." Chemistry of materials 22, 2588-2596 (2010).

[38] A. N. Morozovska, E. A. Eliseev, D. Chen, C. T. Nelson, and S. V. Kalinin. Building Free Energy Functional from Atomically-Resolved Imaging: Atomic Scale Phenomena in La-doped BiFeO3. Phys.Rev. B, **99**, 195440 (2019)

[39] A.Y. Borisevich, E.A. Eliseev, A.N. Morozovska, C.-J. Cheng, J.-Y. Lin, Y.H. Chu, D. Kan, I. Takeuchi, V. Nagarajan, S.V. Kalinin. Atomic-scale evolution of modulated phases at the ferroelectric–antiferroelectric morphotropic phase boundary controlled by flexoelectric interaction. Nature Communications. **3**, 775 (2012)

[40] R. Maran, S. Yasui, E.A. Eliseev, M.D. Glinchuk, A.N. Morozovska, H. Funakubo, I. Takeuchi, N. Valanoor. Interface control of a morphotropic phase boundary in epitaxial samarium-modified bismuth ferrite superlattices. Phys.Rev. **B 90**, 245131 (2014)

[41] D. Chen, C.T. Nelson, X. Zhu, C.R. Serrao, J.D. Clarkson, Z. Wang, Y. Gao, S.-L. Hsu, L.R. Dedon, Z. Chen, D. Yi, H.-J. Liu, D. Zeng, Y.-H. Chu, J. Liu, D.G. Schlom, and R. Ramesh. A Strain-Driven Antiferroelectric-to-Ferroelectric Phase Transition in La-Doped BiFeO3 Thin Films on Si. Nano Lett. **17**, 5823 (2017).

[42] W. Y. Wang, Y. L. Zhu, Y. L. Tang, M. J. Han, Y. J. Wang, X. L. Ma, Atomic Mapping of Structural Distortions in 109 Degrees Domain Patterned BiFeO3 Thin Films, Journal of Materials Research, 32 (12): 2423-2430 (2017).

[43] M. J. Han, Y. J. Wang, D. S. Ma, Y. L. Zhu, Y. L. Tang, Y. Liu, N. B. Zhang, J. Y. Ma, X. L. Ma, Coexistence of Rhombohedral and Orthorhombic Phases in Ultrathin BiFeO3 Films Driven by Interfacial Oxygen Octahedral Coupling, Acta Materialia, 145: 220-226 (2018)

[44] D.C. Arnold, K.S. Knight, G. Catalan, S.A.T. Redfern, James F. Scott, Philip Lightfoot, and Finlay D. Morrison, The β-to-γ Transition in BiFeO₃: A Powder Neutron Diffraction Study, Advanced Functional Materials 20, 2116 (2010).

[45] R. Palai, R. S. Katiyar, Hans Schmid, Paul Tissot, S. J. Clark, Jv Robertson, S. A. T. Redfern, G. A. Catalan, and J. F. Scott, β phase and γ− β metal-insulator transition in multiferroic BiFeO3, Phys. Rev. B 77, 014110 (2008).

[46] Donna C. Arnold, Kevin S. Knight, Gustau Catalan, Simon AT Redfern, James F. Scott, Philip Lightfoot, and Finlay D. Morrison. The β-to-γ Transition in BiFeO3: A Powder Neutron Diffraction Study. Advanced Functional Materials **20**, 2116 (2010).21

**SUPPLEMENTARY MATERIALS**

to

**Melting of Spatially Modulated Phases in La-doped BiFeO$_3$ at Surfaces and Surface-Domain Wall Junctions**

*Anna N. Morozovska[1,2], Eugene A. Eliseev[3], Deyang Chen[4], Vladislav Shvetz[2], Christopher T. Nelson[5], and Sergei V. Kalinin[5,1]*

[1] *Institute of Physics, National Academy of Sciences of Ukraine,*
*46, pr. Nauky, 03028 Kyiv, Ukraine*

[2] *Taras Shevchenko Kyiv National University, Physics Faculty, Kyiv, Ukraine*

[3] *Institute for Problems of Materials Science, National Academy of Sciences of Ukraine,*
*Krjijanovskogo 3, 03142 Kyiv, Ukraine*

[4] *Institute for Advanced Materials and Guangdong Provincial Key Laboratory of Optical Information Materials and Technology, South China Academy of Optoelectronics, South China Normal University, Guangzhou 510006, China*

[5] *The Center for Nanophase Materials Sciences, Oak Ridge National Laboratory,*
*Oak Ridge, TN 37831*

## APPENDIX A. Electrostatic problem

Let us consider a film of thickness *h* placed in a perfect electric contact with conducting bottom electrode that mechanically clamps the film. The top surface of the film is mechanically free and can be in an ideal electric contact with the top electrode, or electrically open, or covered with the surface screening charge. The charge density $\sigma(\varphi)$, appearing due to surface states [1], or electro-chemically active ions [2, 3, 4, 5], depends on the electric potential $\varphi$. Three types of nominally uncharged 180°, 109° and 71° domain walls can exists in BFO and La:BFO.

Electrostatic potential inside the ferroelectric film satisfies the Poisson equation, $\varepsilon_0\varepsilon_b\Delta\varphi - \text{div}\,\vec{P} = 0$ ($\varepsilon_b$ is background permittivity), and Laplace equation is valid in the dielectric gap, i.e. $\varepsilon_0\varepsilon_e\Delta\varphi = 0$ ($\varepsilon_e$ is the dielectric permittivity of external media). Electric boundary conditions are zero electric potential at the bottom of the film contacting the conducting substrate, $\varphi|_{x_2=0} = 0$, and the potential continuity, $\varphi|_{x_2=h-0} - \varphi|_{x_2=h+0} = 0$, at the interface between the ferroelectric film and the ambient medium. Another boundary condition at interface $x_3 = h$ is for the normal components of the electric displacement, namely $D_2|_{x_2=h+0} - D_2|_{x_2=h-0} = \sigma(\varphi)|_{x_2=h}$ where $D_2 = P_2 - \varepsilon_0\varepsilon_b \frac{\partial \varphi}{\partial x_2}$ in the film, ($0 < x_2 < h$) and $D_2 = -\varepsilon_0\varepsilon_e \frac{\partial \varphi}{\partial x_2}$ in the dielectric gap ($h < x_2$). Here, we consider the special case of the surface screening charge with the density given by expression, $\sigma(\varphi) = -\varepsilon_0\varphi/\Lambda$, where $\Lambda$ is the effective screening length [6, 7]. Typically the value of $\Lambda$ is smaller or even significantly smaller than 1 nm [8, 9]. The condition $\Lambda \to 0$ corresponds to the perfect electric contact between the top conducting electrode and the film, and we consider the limit for comparison. The top electrode can be either biased ($\varphi|_{x_2=h} = U$) or grounded ($\varphi|_{x_2=h} = 0$), depending on the experimental situation corresponding to the SPM tip placed on the film surface.

---

[1] Corresponding author, e-mail: sergei2@ornl.gov

Surface energy coefficients $b_i^{(S)}$ and $a_i^{(S)}$ can rule the order parameters behavior as shown in **Fig.S1**.

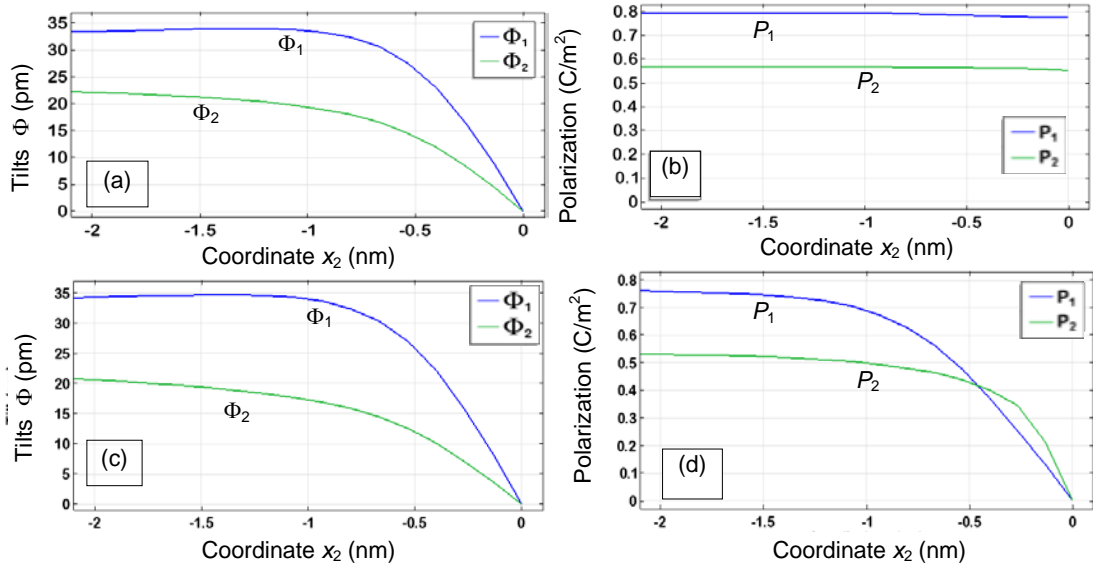

**FIGURE S1**. Depth distribution of the tilts and polarization components at the cross-section $x_1=0$. Plots (a) and (b) correspond to $b_i^{(S)} = 0$ and $a_i^{(S)} = 0$. Plots (c) and (d) correspond $b_i^{(S)} = 0$ and $a_i^{(S)} \to \infty$ (i.e. to $\partial P_i/\partial x_2 = 0$). LGD parameters are listed in **Table III.**

## APPENDIX B. FEM results for pure BiFeO$_3$

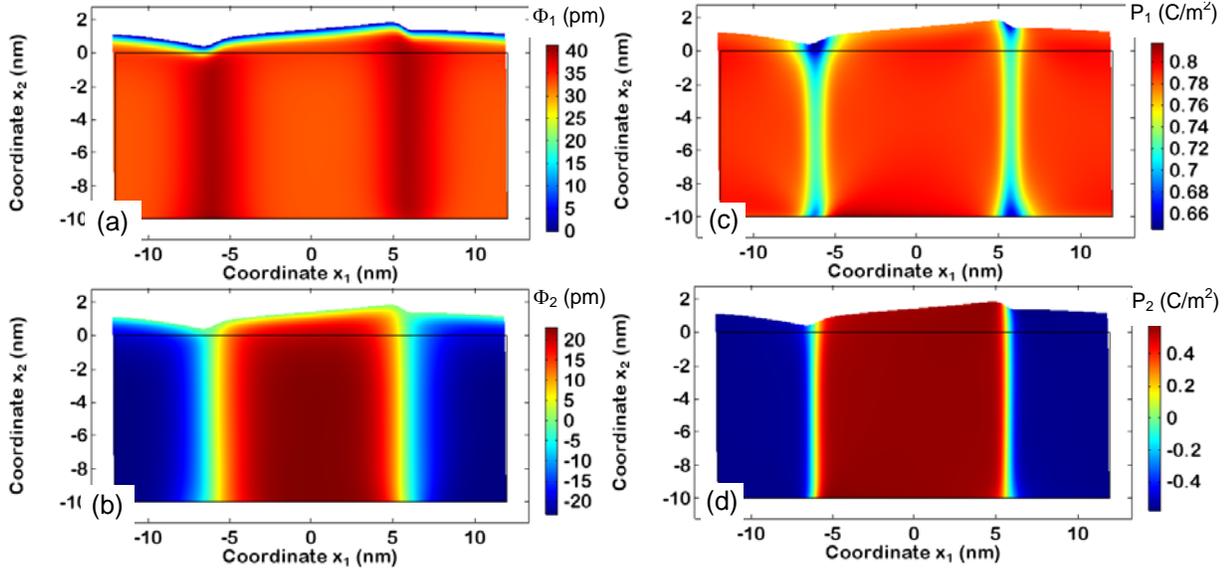

**FIGURE S2**. Distribution of AFD order parameter **Φ (a-b)** and FE polarization **P (c-d)** components, near the twin-wall surface junction in BFO. Here we supposed natural boundary conditions for polarization components and zero tilts at the surfaces. BFO parameters are listed in **Table III.**

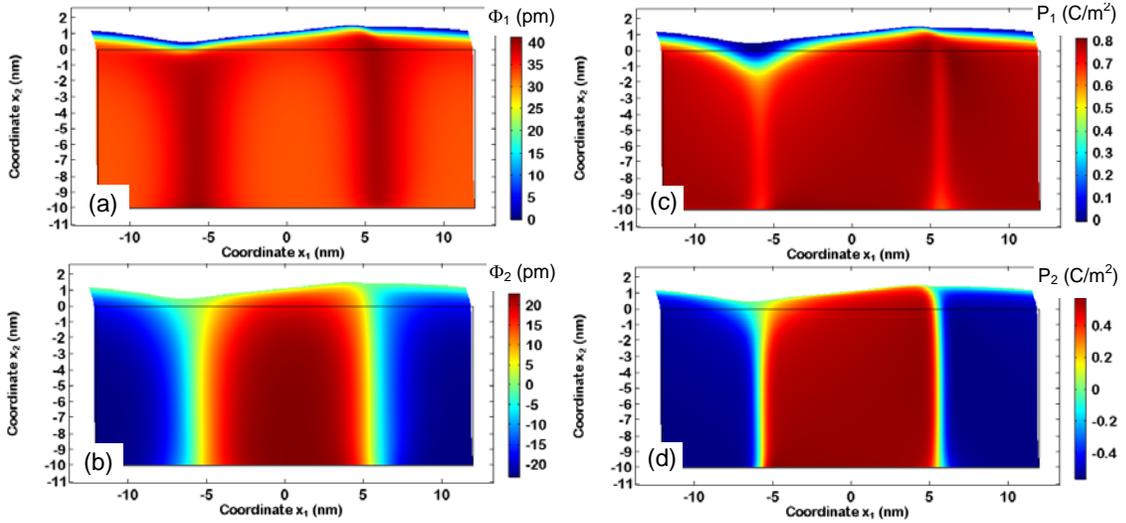

**FIGURE S3**. Distribution of oxygen tilt Φ **(a-b)** and polarization *P* **(c-d)** components near the surface calculated by LGD-approach. Here we supposed zero polarization components and tilts at the surfaces. LGD parameters are listed in **Table III.**